\documentclass[10pt]{article}

\usepackage{amsmath}
\usepackage{amssymb}

\usepackage{cite}

\usepackage{hyperref}

\usepackage{lineno}

\usepackage{microtype}
\DisableLigatures[f]{encoding = *, family = * }

\usepackage{amsfonts}
\usepackage{latexsym}
\usepackage{indentfirst}
\usepackage{color}
\usepackage{lineno}
\usepackage{bm}
\usepackage{url}
\usepackage{amsthm}

\usepackage{graphicx}
\usepackage[dvips]{epsfig}

\usepackage{mathtools}

\usepackage{accents}
\newlength{\dhatheight}

\usepackage[labelfont=bf,labelsep=period,justification=raggedright]{caption}

\makeatletter
\renewcommand{\@biblabel}[1]{\quad#1.}
\makeatother

\date{}

\def\VIZ#1{(\ref{#1})}      
\def\FISHER{\mathcal I}
\def\FISHERR{{\mathcal I}_{\mathcal{H}}}

\def\LAWEXACT{Q_{[0,T]}^{\bm\theta}}

\def\PATHS{\LAWEXACT}

\def\EQUIL{\mu^\theta}

\def\var{{\rm Var}}

\def\EXPECT{\mathbb{E}}
\def\VAR{\operatorname{Var}}
\def\COV{\operatorname{Cov}}
\def\TOL{\operatorname{TOL}}

\def\STATE{\mathbf{x}}

\begin{document}

\begin{flushleft}
{\Large
\textbf{Accelerated Sensitivity Analysis in High-Dimensional Stochastic Reaction Networks}
}
\\
Georgios Arampatzis$^{1}$, 
Markos A. Katsoulakis$^{1,\ast}$, 
Yannis Pantazis$^{1}$
\\
\bf{1} Dep. of Mathematics and Statistics, University of Massachusetts, Amherst, MA, USA \\
$\ast$ Email: markos@math.umass.edu


\end{flushleft}

\section*{Abstract}
Existing sensitivity analysis approaches  are not able to  handle efficiently stochastic  reaction networks
with a large number of parameters and species, which  are  typical in  the  modeling and simulation of
complex biochemical phenomena. 
In this paper,  a two-step strategy for parametric sensitivity analysis for such systems is proposed, exploiting 
advantages and  synergies between  two recently proposed sensitivity analysis methodologies for stochastic
dynamics. The first method performs sensitivity analysis of the stochastic dynamics by means of  the Fisher
Information Matrix on the underlying distribution of the trajectories;  the second method is a
reduced-variance, finite-difference, gradient-type sensitivity approach  relying on  stochastic coupling techniques
for variance reduction. Here we demonstrate that these two methods can be combined  and deployed together
by means of a new  sensitivity bound which incorporates the variance of the quantity of interest as well as the
Fisher Information Matrix estimated from the first method. The first step of the proposed strategy labels
sensitivities using the bound and screens out the insensitive parameters in a controlled manner based also
on the new sensitivity bound. In the second step of the proposed strategy, the finite-difference method is
applied only for the sensitivity estimation of the (potentially) sensitive parameters that have not been screened
out in  the first step. 
Results on an epidermal growth factor network with fifty parameters and on a protein homeostasis with eighty
parameters demonstrate that the proposed strategy is able to quickly discover
and discard the insensitive parameters and in the remaining potentially sensitive parameters it accurately
estimates the sensitivities. The new sensitivity strategy can be several times faster than  current state-of-the-art
approaches that test all parameters, especially in ``sloppy'' systems. In particular, the computational acceleration
is quantified by the ratio between the total number of parameters over the number of the sensitive parameters.

\section*{Author Summary}
Modeling and simulation of complex phenomena in systems biology has been established in recent years as an
important scientific research tool that can lead to a profound understanding of fundamental biological mechanisms.
Such mathematical models   typically  rely  on large stochastic biochemical reaction networks with  a large number
of parameters associated with individual reactions.
Sensitivity Analysis quantifies the response of the system to perturbations and  is  an indispensable tool for
the modeling and parametrization of complex biochemical reaction networks.  However, sensitivity analysis for
stochastic systems is generally a challenging  problem primarily due to the high variance of the estimated quantities
of interest, as well as the very large number of parameters that typically appear in realistic complex biochemical
networks. In this paper, we address these challenges with a proposed  two-step sensitivity analysis strategy. 
Concepts from estimation theory as well as from information theory are utilized in the first step of our
strategy in order to screen out insensitive parameters, and thus reduce computational cost in a systematic and
controllable fashion, while a finite-difference algorithm is deployed in the second step, focusing
only on the remaining  sensitive parameters. We demonstrate that the proposed strategy is capable of handling
stochastic reaction networks with hundreds of parameters (and potentially many more) in a hierarchical and
efficient manner.

\section*{Introduction}
\label{intro:sec}

Biological and biochemical reaction networks provide a powerful computational and modeling
tool for the intrinsic understanding of fundamental mechanisms in systems biology such as
metabolic, regulatory and signaling pathways. With the advent of ever-increasing
computational power and the desire for more accurate representations of the physical processes
at the level of the (sub-)cell or at the level of populations, larger, more complex and
more sophisticated biochemical reaction networks have been developed. For instance,
reconstruction for genome-scale, steady-state models of metabolic networks, macromolecular
synthesis and multiscale systems biology necessitates reaction networks with (tens of) thousands
of reactions \cite{Bader:07,Henry:10,Theile:10}. The enormous increase in the size of modeled reaction networks
 presents severe modeling and computational challenges, among others  the understanding, designing, inferring and predicting
the properties and the behavior of the output, especially when the stochasticity is a prerequisite
for  correct modeling, e.g. \cite{McAdams:97,Thattai:01}. Indeed, for low  species populations, stochastic
models are critical for the correct representation of the inherent randomness and discrete
nature of intracellular networks.
In particular, a ubiquitous problem in biochemical reaction networks is to quantify the response of the
system when perturbations on the input or on the parameters of the system are performed.
Subsequently, questions on the robustness, (structural) identifiability,
experimental design, uncertainty quantification, estimation and  control  can be addressed, \cite{Saltelli:08}. 
The quantification of  system response to parameter perturbations
is called sensitivity analysis and  is an indispensable analysis tool for the study of kinetic
models \cite{Saltelli:08,Distefano:13}. 

For large-scale reaction networks, sensitivity analysis is
especially important  due to the  confluence of the  nonlinear,
stochastic and typically  non-equilibrium statistical mechanics characteristics of the  models. Additionally,
the large number of parameters yields a very high number of sensitivity indices needed to be estimated,
increasing by orders of magnitude  the overall computational cost 
when compared to simply the forward simulation of the model. Finally, the stochasticity may also result in high variance in the estimators 
of the sensitivity indices, adding both  further computational  cost and uncertainty in the predictions of the sensitivity analysis.
This
paper addresses precisely such challenges, namely  the parametric sensitivity analysis of high-dimensional stochastic reaction networks, both in the size of the
parameter vector (large parameter space) and the number of species (large state space).

Recently, there has been significant progress in developing sensitivity analysis tools for
low-dimensional stochastic dynamics, modeling well-mixed chemical reactions and biological
networks. Some of the mathematical tools includes log-likelihood methods and Girsanov
transformations \cite{Glynn:90,Nakayama:94,Plyasunov:07}, polynomial chaos expansions \cite{Kim:07},
finite difference methods and their variants \cite{Rathinam:10, Anderson:12,AK:2013}, as well as pathwise
sensitivity methods \cite{Khammash:12}. 
%
In another
direction, recent sensitivity analysis approaches have been proposed as means to  quantify the overall
behavior of the reaction network and not just the response of a specific observable function.
These sensitivity analysis methods employ information theory metrics such as the relative
entropy (also known as Kullback-Leibler divergence)
as well as the Fisher Information Matrix (FIM). Moreover, taking into account the fact that the
knowledge of the stationary distribution is rarely known in biochemical reaction networks,
these information-based methods resort either on linearized Gaussian approximations of the underlying process
\cite{Komorowski:11} or they rely on  path-space distribution calculations, \cite{Pantazis:Kats:13,PKV:2013}.
The latter approach is exact since no approximation is necessary, 
while  it is  gradient-free in the sense that a single model (parameter)  simulation is carried out, resulting in reduced-variance estimators.
Overall, gradient-free sensitivity analysis methods such as the ones proposed in \cite{Komorowski:11,Pantazis:Kats:13,PKV:2013} are highly appropriate for systems with a high-dimensional parameter space since they allow for an efficient exploration of the parameter space without the calculation of a very high number of directional derivatives.


However, existing sensitivity analysis methods are not  capable of estimating the sensitivity analysis of 
specific quantities of interest (observables)  for large stochastic networks, in a computationally efficient and accurate manner.
Indeed,
beginning with the deterministic models, their sensitivity analysis is fairly easily  formulated by the
adjunct system of differential equations but this  is not always adequate because such models  do not take
into account the intrinsic stochasticity which crucially affects the behavior of the system \cite{McAdams:97,Thattai:01}.
On the other hand, stochastic sensitivity analysis methods based on information criteria \cite{Komorowski:11,Pantazis:Kats:13}
are also inexpensive  in terms of computational cost since they can handle large networks with many parameters due to their gradient-free nature,
\cite{PKV:2013}, however, they are not observable-based, hence may not provide precise analysis for specific  quantities of interest.
Finally, observable-based approaches such as finite-difference (gradient) methods
can have an overwhelming computational cost, either due to high variance in the gradient
estimators 
\cite{Vlachos:12} or the
high-dimensional state space \cite{AK:2013}; in the latter case such computations can be prohibitively expensive
when   they need to be run over and over again for many parameter perturbations  related to the 
sensitivity indices. 



Here we demonstrate that the aforementioned methods can be combined  and deployed together by means of a new  sensitivity
bound which incorporates the variance of the quantity of interest as well as the Fisher
Information Matrix, 
see inequality (\ref{sens:bound:gen}). 
The proposed strategy is a two-step hierarchical approach where in the first step the insensitive
observables and parameters are found and eliminated from further analysis with controlled accuracy;  the second step targets
the remaining (potentially) sensitive parameters and observables:
\begin{itemize}
\item {\em Step 1}: Screen the parameters through a computational inexpensive labelling of the insensitive
parameters based on a sensitivity bound (SB) derived from a path-space   Cramer-Rao inequality (see \VIZ{sens:bound:gen}). The sensitivity index (SI)
for an observable (see \VIZ{sens:index} for a definition) can be bounded by the square root of the variance
of the observable  multiplied by the diagonal FIM elements.
Here we utilize the pathwise FIM \cite{PKV:2013} which quantifies information  from both the steady
states and  the dynamics of the reaction network.
\end{itemize}
To this end  note that since the SB is an upper bound of the SIs (see \VIZ{sens:bound:gen}), neither guarantees that large values of the bound imply
large SIs,  nor infers information on their order. Therefore, for the SIs where the SB is large,  a more
accurate sensitivity analysis method is needed. Indeed, in   the second step of the strategy, an observable-based
sensitivity analysis algorithms is applied to the potentially sensitive parameters and observables:
\begin{itemize}
\item {\em Step 2}: Employ an estimator for  the SIs on  the remaining
potentially sensitive parameters. Here we choose the gradient estimator given by the coupling method \cite{Anderson:12}
which is a finite-difference approach with reduced variance, even in high-dimensional systems, \cite{AK:2013}.
\end{itemize}

The pathwise FIM quantifies the information change of the path-space distribution  (i.e., the distribution of the species trajectories) of stochastic
processes under perturbations, \cite{PKV:2013}.
Furthermore, the estimation of the
pathwise FIM is very efficient because  a single model simulation is required for the
computation of the whole matrix, while  the variance of the statistical estimator is typically small.
%
The acceleration in sensitivity analysis due to  the proposed strategy can be very significant  especially when sloppy systems
are considered,  \cite{Gutenkunst:07,Erguler:11,PKV:2013},  and most of the parameters are expected to  be screened out as insensitive from {\em Step 1}. Moreover, the
proposed strategy offers a simple way to rationally balance  accuracy and computational cost, selecting 
the number of  insensitive parameters that need to be discarded.
The discrimination between sensitive and insensitive parameters can be performed by a user-determined
tolerance or by the availability of computational resources. The proposed strategy, through the SB, guarantees that the SIs for the insensitive parameters
will lie below the value of the tolerance. 

A detailed demonstration and validation of the proposed strategy applied on three biological models is provided.
The p53 model  \cite{Prives:98,Harris:05, Geva-Zatorsky:06} is presented as a small-size example,
having 21 SIs, introducing the idea of screening out negligible sensitivities. Due to the nontrivial stationary
behavior of the p53 model, with random and persistent oscillations of the solution, sensitivities of observables
such as the amplitude and the frequency of the oscillations are explored. A more complex model with
a large number of species and parameters, the Epidermal Growth Factor Receptor (EGFR) model \cite{Moghal:99, Hackel:99,Schoeberl:02},
is studied  next. The EGFR model is studied both in the transient and in the stationary regimes allowing
us to demonstrate the applicability of the proposed strategy in both settings. In this particular example,
more than half of the sensitivities can be excluded by the computationally inexpensive SB of {\em Step 1}.
Finally, a protein homeostasis model, \cite{Proctor:2011},  with a total of  $4160$ SIs is presented as a "sloppy"
example,  where more that $85\%$ of the total SIs can be safely ignored having  a guaranteed maximum bound  by the  screening
in {\em Step 1}. The results of the proposed strategy are compared against the results of the 
full coupling method applied to all SIs without any  screening. 
The computational cost of the two methods, measured in number of samples, is compared
and found that the two-step strategy can accelerate up to approximately $\frac{K}{K'}$ times the sensitivity
analysis where $K$ is the total number of parameter, while  $K'$ is the number of parameters remaining after the screening in {\em Step 1};
as we show in  the demonstrated examples, the SB calculation in {\em Step 1} is negligible since it is at least
one order of magnitude less expensive compared to a single run of the coupling method in {\em Step 2}.

The paper is organized as follows. In the Methods section, the sensitivity analysis strategy is presented.
The
validation of the proposed sensitivity analysis approach is provided in the Results section, where one small and two
large biochemical reaction networks are  tested while the computational advantages are  quantified
and presented in the Discussion section. 

\section*{Methods}

A well-mixed reaction network with $N$ species, $\mathbf{S} = \{S_1,\ldots,S_N\}$,
and $J$ reactions, $\mathbf{R} = \{R_1,\ldots,R_J\}$ is considered. The state of the system at any time
$t\geq0$ is denoted by an $N$-dimensional vector $\mathbf{X}_t = [X_{t,1},\ldots,X_{t,N}]^T$ where
$X_{t,i}$ is the number of molecules of species $S_i$ at time $t$. Let the $N$-dimensional
vector $\mathbf{\nu_j}$ correspond to the stoichiometry vector of $j$-th reaction such that $ \nu_{i,j}$
is the stoichiometric coefficient of species $S_i$ in reaction $R_j$. Given that the reaction
network at time $t$ is in state $\mathbf{X}_t = \mathbf{x}$, the propensity function, $a_j^\theta(\mathbf{x})$,
is defined so that the infinitesimal quantity $a_j^\theta(\mathbf{x})dt$ gives the probability 
of the $j$-th reaction to occur in the time interval $[t,t+dt]$. Propensities are typically
dependent on the state, $\mathbf{x}$, of the system and the reaction conditions of the network
which are made explicit by the parameter vector $\theta\in\mathbb R^K$.
Mathematically, $\{\mathbf{X}_t\}_{t\in\mathbb R_+}$ is a continuous-time
Markov chain (CTMC) with countable state space $E\subset \mathbb N^N$. The transition
rates of the CTMC are the propensity functions $a_j^\theta(\cdot),\ j=1,\ldots,J$.
The transition rates determine the clock of the updates (or jumps) from a current state
$\mathbf{x}$ to a new (random) state $\mathbf{x}'=\mathbf{x} + \mathbf{\nu_{j}}$ through the total rate
$a_0^\theta(\mathbf{x}):=\sum_{j=1}^J a_j^\theta(\mathbf{x})$ while the transition probabilities
of the process are determined by the ratio $\frac{a_j^\theta(\mathbf{x})}{a_0^\theta(\mathbf{x})}$.
In order to have a complete description of the reaction network, an initial distribution
of the state at time instant $t=0$ denoted by $\nu^\theta(\cdot)$ is also needed.

There are exact algorithms for the simulation of the reaction network such as the
stochastic simulation algorithm (SSA) of Gillespie \cite{Gillespie:76, Gillespie:77} 
or the next reaction algorithm of Gibson and Bruck \cite{Gibson:00} or the constant-time
kMC algorithm of Slepoy et al. \cite{Slepoy:08} as well as inexact
approximation algorithms such as $\tau$-leap \cite{Gillespie:01} and several
variations of it \cite{Rathinam:03, Chatterjee:05}. As an example, given that the
system is at the state $\mathbf{X}_t = \mathbf{x}$ at time $t$, SSA computes the waiting
time $\delta t$ as a random number drawn from an exponential distribution with
the total rate $a_0^\theta(\mathbf{x})$,  while the $R_{j^*}$ reaction
occurs when $j^*\in\{1,\ldots,J\}$ is chosen such that
$\sum_{j=1}^{j^*-1}a_j^\theta(\mathbf{x}) < u a_0^\theta(\mathbf{x}) < \sum_{j=j^*}^{J}a_j^\theta(\mathbf{x})$
where $u$ is a random number uniformly chosen in the interval $[0,1]$. The
new state is given by $\mathbf{X}(t+\delta t) = \mathbf{x}' = \mathbf{x} + v_{j^*}$.

The path space distribution of the stochastic process on the time-interval $[0,T]$ is denoted by
$\PATHS$. Notice that the dependence of the path space distribution to the initial distribution
is made implicit for notational simplicity. Intuitively, the path space is the set of all possible trajectories in $[0, T]$, 
generated by SSA for the particular reaction network while the path space distribution is the
probability to observe a particular trajectory. For a concrete and simple  example of a path distribution, we refer
to supporting information File S1 (Sec. 2) 
where Discrete-Time Markov Chains DTMCs are considered.
As we shall show next (see also File S1), 
the path space perspective is easy to implement exploiting concepts from information theory and
from non-equilibrium statistical mechanics.

We now turn our attention to observables of the stochastic process, $\mathbf{X}_t$. We denote by
$\mathbf{F}(\cdot) = [F_1(\cdot),\ldots,F_L(\cdot)]^T$ the vector with $L$ state-dependent
observable functions, $F_\ell:\mathcal X \rightarrow \mathbb R,\ \ell=1,\ldots,L$. Two typical options for the
observable function are the time-average of a function as well as the value of a function at a specific time
instant. The time-average observable is defined in a general setting as
\begin{equation}\label{time:average}
F_\ell(\{\mathbf{X}_t\}_{0}^T) = \frac{1}{T}\int_{t=0}^T f_\ell(\mathbf{X}_t) dt \ ,
\end{equation}
while the time-specific observable is defined as 
\begin{equation}\label{final:time}
F_l(\{\mathbf{X}_t\}_{t=0}^T) = f_l(\mathbf{X}_T) \ .
\end{equation}
The most common observable is
the population of the $\ell$-th species, i.e., the projection of the state vector to the $\ell$-th direction
($f_\ell(\mathbf{x})=x_\ell$), however, other observable functions such as correlations
between various species of the network, time-correlations for a specific species as well
as switching or exit times can be considered. Another important discrimination for the
observable functions stems from the time regime where the stochastic process is sampled.
There are two important regimes; the stationary regime where the process is at equilibrium
and the transient regime where the stochastic process initialized far from equilibrium and
in the course of time it relaxes towards the steady states. At the stationary regime, the initial
distribution of the stochastic process is the stationary distribution and
both time-average and time-specific observables produce the same ensemble averages
since  $\EXPECT_{\PATHS}[\frac{1}{T}\int f(\mathbf{X}_t) dt]
= \EXPECT_{\PATHS}[f(\mathbf{X}_T)] = \EXPECT_{\EQUIL}[f(\mathbf{x})]$
where $\EQUIL:E\rightarrow\mathbb R$ is the stationary distribution while $\EXPECT_P[f]$
denotes the expectation of $f(\cdot)$ with respect to the probability $P$ (i.e., $\EXPECT_P[f(x)] := \int f(x)P(x)dx$).
Moreover, due to the (assumed) ergodicity property of the reaction network, it is typical
to obtain ensemble averages and statistics  from time-averaged observables
since ergodicity asserts that
$\lim_{T\rightarrow\infty}\frac{1}{T}\int f(\mathbf{X}_t) dt = \EXPECT_{\mu^\theta}[f(\mathbf{x})]$. 

The goal of this paper is to describe an efficient and highly resolved strategy to compute the
parameter sensitivities on the observable functions, i.e., to compute the sensitivity matrix,
$S\in\mathbb R^{K\times L}$, defined element-wise by
\begin{equation}
S_{k,\ell} = \frac{\partial}{\partial\theta_k} \EXPECT_{\PATHS}[F_\ell(\{\mathbf{X}_t\}_{t=0}^T)] \ , \ \ \ k=1,\ldots,K \ \&\ \ell=1,\ldots,L \ .
\label{sens:index}
\end{equation}
The element, $S_{k,\ell}$, is the Sensitivity Index (SI) of the $\ell$-th observable to the $k$-th parameter.
The proposed strategy is separated into two steps where, in the first step, a Sensitivity Bound (SB) is computed for each
SI. The evaluation of the  SB is based on tools from estimation theory~\cite{Casella2002,Kay:93}
and information theory~\cite{Cover:91}. The SB is a product of two factors where the first one  depends
on the properties of the observable function while the second factor depends only on the properties
of the underlying path space distribution of the stochastic process (see \VIZ{sens:bound:gen} and  \VIZ{sens:bound:statio}
below). The computational efficiency of the SB stems from its factorization
into two terms each one quantifying different aspects of the SIs. Then, in the second
step, we apply a  computationally more expensive but accurate sensitivity estimation method. In particular, we use  the coupling method~\cite{Anderson:12},
however,  applied  only on the potentially sensitive SIs since from the first step the least sensitive SIs have
been screened out with a controlled error given by the SB. We discuss these two components of our proposed methodology next.

\subsection*{{Step 1}: Screening out  insensitive parameters and observables}
In parameter estimation theory, the Cramer-Rao theorem states that
the variance of an estimator cannot be smaller than the inverse of the FIM \cite{Casella2002, Kay:93}.
Here, we are not interested to  bound the variance of the estimator from below as in the Cramer-Rao
theorem, but rather  to bound  the bias of the estimator from above which in our context corresponds to the  SI. Thus,
the sensitivity bound (SB) can be obtained by rearranging the generalized Cramer-Rao bound for a biased estimator, \cite{Casella2002, Kay:93}.
In particular, when  the distribution in the Cramer-Rao theorem is the path distribution, the absolute SI of the $\ell$-th
observable is bounded by the inequality
\begin{equation}
|S_{k,\ell}| \le  B_{k,\ell} :=  \sqrt{\var_{\PATHS}(F_\ell)} \sqrt{\FISHER(\PATHS)_{k,k}} \ ,
\label{sens:bound:gen}
\end{equation}
where $\FISHER(\PATHS)$ is the $K\times K$ pathwise FIM. We also recall that the  path space distribution of the stochastic process on the time-interval $[0,T]$ is denoted by $\PATHS$. 
This inequality is a general SB which, assuming that the estimation of the variance of the observable
and the pathwise FIM is tractable and fast, can be utilized to discard the most insensitive SIs. Indeed,
if the right hand side of the inequality is small then the corresponding SI is also small.
In other words, given a specific observable, diagonal FIM elements with small values imply
low SIs. However, notice that  large FIM values in \VIZ{sens:bound:gen} do not necessarily imply large
SIs or any information on ranking the SIs with high values. From this latter observation stems the need for  the second step
in the proposed sensitivity analysis strategy.

Overall, with the cost of estimating an upper bound instead of the actual values,
the estimation of $K\times L$ sensitivity indices is reduced to the estimation
of $L$ variances and $K$ elements of the FIM which is a significant  reduction especially
when the studied system is high-dimensional both in the parameter space ($K\gg1$) and
in state space ($N\gg1$). In typical cases, $F_\ell(\mathbf{x})=x_\ell$ (i.e., projection operators as
observables) hence $L=N\gg1$, however, when correlations between species are of
interest then $L=N(N-1)/2$ and the computation of the sensitivity matrix \VIZ{sens:index} becomes readily
intractable.

From an information theory perspective, pathwise FIM is the Hessian of the pathwise relative
entropy which geometrically corresponds to the curvature around its minimum value \cite{Pantazis:Kats:13}. For a
definition of pathwise relative entropy as well as its properties for both discrete-time Markov chain (DTMC) and CTMC cases
we refer to supporting information File S1. 
The SB (eq. \VIZ{sens:bound:gen}) can be also derived as a limit of a variational inequality
which relates the weak error between two path distributions with their pathwise relative entropy (see File S1 for details). 
Another derivation of the SB can be obtained as a limit of the Chapman-Robbins inequality which
incorporates the chi-squared divergence \cite{Casella:point}.


We also remark that \VIZ{sens:bound:gen} can be generalized to
provide a SB for any combination of the parameters (i.e., bound the directional
derivative).
Indeed, 
for any $v\in\mathbb R^K$ with $|v|=1$ and denoting the directional
derivative by $\partial_v \EXPECT_{\PATHS}[F]$, it holds that  the sensitivity
bound for an  observable function $F(\cdot)$ at the direction $v$ is given by
\begin{equation}
|\partial_v \EXPECT_{\PATHS}[F]| \le \sqrt{\var_{\PATHS}(F)} \sqrt{v^T\FISHER(\PATHS) v} \ .
\label{sens:bound:direc:deriv}
\end{equation}
Furthermore, inequality \VIZ{sens:bound:direc:deriv} combined with the spectral analysis of pathwise
FIM can infer the least sensitive directions of the system as well as the most sensitive candidate
directions \cite{PKV:2013,Komorowski:11}.

The computation of the pathwise FIM, $\FISHER(\PATHS)$, necessitates the explicit knowledge of the probability
function which is not always possible. However, in the setting of Markov processes, explicit formulas
for the pathwise FIM exist \cite{PKV:2013}. Indeed, using the properties for the pathwise relative
entropy, we are able to derive explicit formulas for the pathwise FIM.
Next, we provide such formulas for both the stationary and the transient regime.
Note that, as expected,  the SB is time-independent  in the stationary regime.

\subsubsection*{Stationary regime} \label{sec:stationary:regime}
In  stationary regimes of stochastic processes,  the probability law of the process is time-independent. Therefore, the typical
observables utilized in  steady state regimes  are the time-averaged observables given by
\VIZ{time:average}. Then, it can be shown  for the stationary regime and time-averaged
observables that the SB given by \VIZ{sens:bound:gen} becomes time-independent and it can be rewritten  as  
\begin{equation}
|S_{k,\ell}| \le \sqrt{\tau_{\EQUIL}(f_\ell)} \sqrt{\FISHERR(Q^{\theta})_{k,k}} \ ,
\label{sens:bound:statio}
\end{equation}
where $\tau_{\EQUIL}(f_\ell)$ is the Integrated Autocorrelation Time (IAT) while $\FISHERR(Q^{\theta})$
is the FIM of the so called Relative Entropy Rate (RER) \cite{Pantazis:Kats:13,PKV:2013} (for a proof, see File S1, Sec. 1.1).
The IAT is given by (see File S1, Sec. 3.2 for a derivation) 
\begin{equation}
\tau_{\EQUIL}(f) = \int_{-\infty}^{\infty} <f(\mathbf{X}_t)-\EXPECT_{\EQUIL}[f], f(\mathbf{X}_0)-\EXPECT_{\EQUIL}[f]>_{\EQUIL} dt
\label{IAT:CTMC}
\end{equation}
where $<f(\mathbf{X}_t)-\EXPECT_{\EQUIL}[f(\mathbf{x})], f(\mathbf{X}_0)-\EXPECT_{\EQUIL}[f(\mathbf{x})]>_{\EQUIL}$
is the stationary covariance between $f(\mathbf{X}_t)$ and $f(\mathbf{X}_0)$.
We remark that IAT has been used in a series of problems in probability and statistics
to measure the performance of samplers and estimators, \cite{junliu}. Many of the state-of-the-art
estimators of IAT give non-reliable results while those that can give satisfactory results depend
on parameters that need to be tuned by visual inspection of the autocorrelation time.
In fact, we have found   that the most reliable estimator of IAT is the one presented in File S2 of 
the supporting information.

In the context of well-mixed reaction networks and assuming smoothness of the propensity functions,
$a_j^{\theta}(\mathbf{x}), \ j=1,\ldots,J$, with respect to the parameter vector, $\theta$, the pathwise FIM is
explicitly written as, \cite{PKV:2013},
\begin{equation}
\FISHERR(Q^{\theta}) = \EXPECT_{\EQUIL}\Big[ \sum_{j=1}^J a_j^\theta(\mathbf{x})
\nabla_\theta \log a_j^\theta(\mathbf{x}) \nabla_\theta \log a_j^\theta(\mathbf{x})^T \Big] \ .
\label{FIM:CTMC}
\end{equation}
Thus, the time-independent pathwise FIM  can be practically estimated as
an ergodic average. Statistical estimators for the stationary pathwise FIM are  provided
in the File S2. 

\smallskip
\noindent
{\it Remark:} As a concrete example of (\ref{sens:bound:statio}) consider a stochastic process whose autocorrelation function
with respect to an observable $f$ decays exponentially fast with rate (i.e., decorrelation time) $\tau_d=\tau_d(f,\EQUIL)$.
Then, the  IAT satisfies  $\tau_{\EQUIL}(f) = 2\tau_d\var_{\EQUIL}(f)$, \cite{junliu},  and then the  stationary SB (\ref{sens:bound:statio}) becomes
\begin{equation}
|S_{k,\ell}| \le \sqrt{2\tau_d} \sqrt{\var_{\EQUIL}(f_\ell)} \sqrt{\FISHERR(Q^{\theta})_{k,k}} \ .
\label{sens:bound:example}
\end{equation}

\subsubsection*{Transient regime}
In the transient regime, 
the sensitivity bound is given by the general inequality \VIZ{sens:bound:gen}.
Nevertheless, in terms of implementation, the estimators of the transient regime are the same
as in the stationary regime. This is evident in  \VIZ{FIM:trans} below,  which is derived
in File S1, Sec. 3.1. 

More precisely, in order to derive the explicit formula for the pathwise FIM, we slightly adapt the
notation for the propensity functions. We define the transition rate function as
\begin{equation*}
a^\theta(\mathbf{X}_{t-},\mathbf{X}_t) = \left\{\begin{array}{cl} a^\theta_j(\mathbf{X}_{t-}) & \text{if } \mathbf{X}_{t} = \mathbf{X}_{t-} + \mathbf{\nu}_j\\
0 & \text{otherwise} \end{array} \right.\ .
\end{equation*}
In the new notation, $\mathbf{X}_{t-}$ denotes
the state of the process just before the jump time $t$ while $\mathbf{X}_{t}$ denotes the new state of
the process after the jump. Then, assuming that the propensity functions are differentiable with respect to 
 the parameter vector $\theta$, the pathwise FIM for the transient regime is given by
\begin{equation}
\FISHER(\PATHS) = \FISHER(\nu^{\theta})  + \int_{0}^{T} \FISHERR(Q_{t}^\theta) dt\ ,
\label{FIM:trans}
\end{equation}
where $\FISHER(\nu^{\theta})$ is the FIM of the initial distribution, $\nu^\theta$, while
$\FISHERR(Q_{t}^\theta)$ can be viewed as the instantaneous pathwise FIM given by
\begin{equation}
\FISHERR(Q_{t}^\theta) = \mathbb E_{Q_{[0,t]}^{\theta}}\left[ a_0^{\theta}(\mathbf{X}_{t-})
\nabla_\theta \log a^\theta(\mathbf{X}_{t-}, \mathbf{X}_{t}) \nabla_\theta \log a^{\theta}(\mathbf{X}_{t-}, \mathbf{X}_{t})^T \right] \ ,
\end{equation}
which readily reduces to \VIZ{FIM:CTMC}, in the stationary regime.
The formula of the instantaneous pathwise FIM is not as simple as in the DTMC case (see File S1, Sec. 2.1) 
because the waiting time of the jumps is now random, however, the statistical estimator for the pathwise
FIM in the transient regime is as simple as in the stationary regime. In fact it has exactly the same formula
as we show in File S2. 

\subsection*{{ Step 2}: Finding and ranking the most sensitive SIs}
In this second
step of the proposed strategy, we employ a  computationally more expensive but accurate sensitivity
estimation method: we use the  the coupling method~\cite{Anderson:12}, which is only applied   on the
potentially sensitive SIs since from the {\em Step 1}  the least sensitive SIs have
been screened out with a controlled error given by the SB. We discuss the coupling  methodology next.

First, the SI defined in (\ref{sens:index}) can be approximated by a second-order finite difference scheme as
\begin{equation}
S_{k,\ell} \approx \tilde S_{k,\ell} = \frac{1}{2\epsilon_0} \Big(\EXPECT\big [F_\ell (\mathbf{X}^+ )\big ]
- \EXPECT\big[F_\ell (\mathbf{X}^- )\big] \Big) \ ,
\label{derivative:estimator}
\end{equation}
where we use the abbreviated notation $\mathbf{X}^\pm = \{\mathbf{X}_{t}^{\theta_k\pm\epsilon_0}\}_{t=0}^T$ while
$\epsilon_0\in\mathbb R$ and $\epsilon_0\ll1$. In this study we set $\epsilon_0=0.1$. Notice also that, for notational
simplicity, we dropped the dependence on the underlying path space distribution from the expectation.
The variance of the estimator in (\ref{derivative:estimator}) is proportional to 
\begin{equation}
\begin{aligned}
	\VAR\big[ F_\ell(\mathbf{X}^+) -  F_\ell(\mathbf{X}^-) \big] 
	&= \VAR\big[ F_\ell^2(\mathbf{X}^+)\big] + \VAR\big[ F_\ell^2(\mathbf{X}^-)\big] - \COV \bigl ( F_\ell(\mathbf{X}^+),F_\ell(\mathbf{X}^-)\bigr) \ .
\label{var:estimator}
\end{aligned}
\end{equation}
In order to minimize the variance of the estimator we have to correlate the processes in a way such that the covariance
$\COV \bigl ( F_\ell(\mathbf{X}^+),F_\ell(\mathbf{X}^-)\bigr)$
is maximized since the first two terms in (\ref{var:estimator}) do not depend on the correlation
between the two processes, $\mathbf{X}^+$ and $\mathbf{X}^+$.  One way to correlate these two processes is the stochastic coupling method, 
\cite{ Anderson:12, AK:2013}, where it has been proved that the coupling between the two processes
indeed reduces the variance of the estimator of \VIZ{derivative:estimator}.  File S3 
in the supporting information contains implementation details regarding the coupling method.
Notice also that the coupling method is employed $K$ times; one for each parameter, $\theta_k, \ k=1,...,K$.

\begin{figure}
\centering
\includegraphics[width=0.99\textwidth]{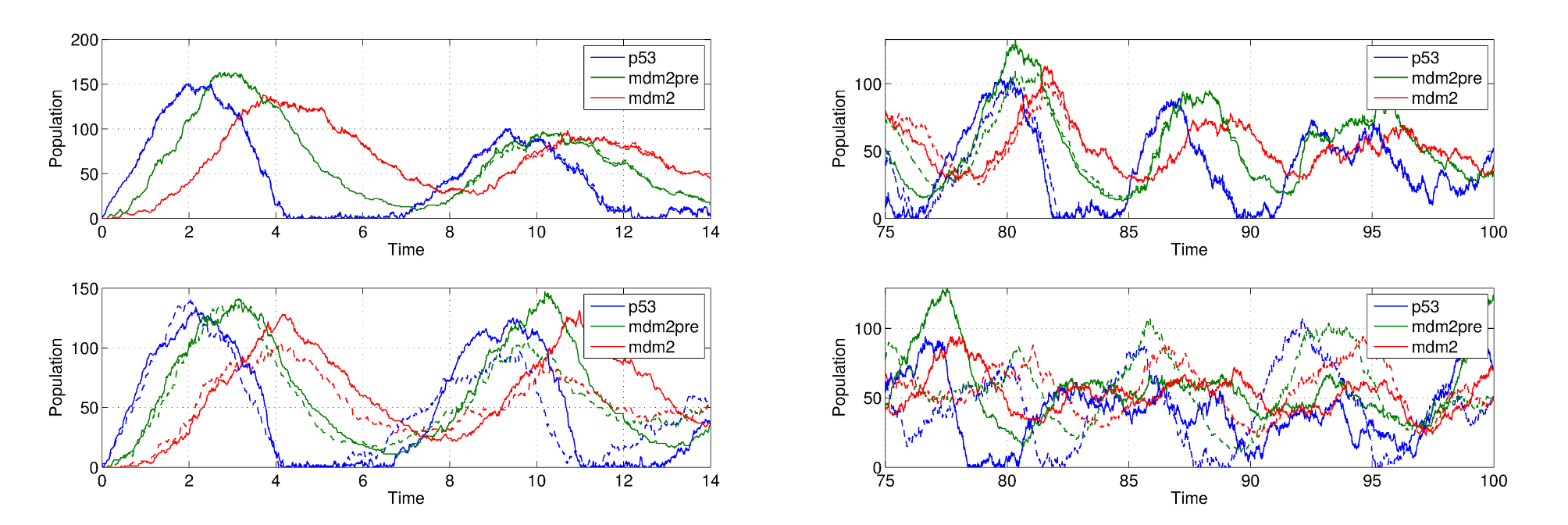}
\caption{ Trajectories of the species of the p53 model. The solid and the dashed lines correspond to
the unperturbed and the perturbed parameters respectively. Left panel: The result of the coupled algorithm (upper
plot) in comparison to the result of two completely uncoupled runs (lower plot) is presented. Note that the
coupled algorithm produces correlated paths, that are close to each other, thus leading to smaller variance
in the estimator (\ref{derivative:estimator}). Right panel: The same computation as in the left panel for a larger time interval.
The coupling method manages to keep the trajectories very close.}
\label{fig:trajectories}
\end{figure}

In Figure~\ref{fig:trajectories}, the trajectories of the species of the p53 model (details in Results
below) obtained from the coupling method are compared with two completely uncoupled trajectories.
Even when the processes have random oscillations, the coupling method manages to keep the trajectories very
close. As it is also evident from  Figure~\ref{fig:trajectories} this is also true even for long times. On the other
hand, the uncoupled trajectories start to separate shortly after their starting point and for longer times the peaks of the
oscillatory trajectories are in completely different positions as it can be seen in the right lower plot of
Figure~\ref{fig:trajectories}.

\subsection*{How to use the proposed strategy}
In this subsection, we describe some of the ways  the proposed methodology can be used in practice,
but clearly other approaches are also possible.

\vspace{5pt}
\noindent\textbf{Setting a maximum number of SIs.}
In this first perspective, we assume that the user  is given a computational resources budget that allows
the simulation of at most $M$ sensitivity indices. Two important questions arise:
\begin{enumerate}
\item[(a)] Which SIs, $S_{k,\ell}$, should be chosen?
\item[(b)] Are there any guarantees regarding the magnitude of the remaining SIs?
\end{enumerate}
In order to answer these questions we sort the SBs, see (\ref{sens:bound:gen}), in a linear descending fashion   and define the set
\begin{equation}\label{CM:set}
C_M := \left \{ \textrm{ all }    (k,\ell): \;  B_{k,\ell} \textrm{ is one of the } M \textrm{ largest SBs } \right \}.
\end{equation}
Let also
\begin{equation}
\TOL(M) := \max_{(k,\ell) \notin C_M}B_{k,\ell}.
\end{equation}
Then the answers to the above questions are: (a) Choose all pairs $(k,\ell)\in C_M$, (b) for all $(k,\ell) \notin C_M$ we have from the inequality (\ref{sens:bound:gen}) that,
	\begin{equation} \label{bound:TOL}
		|S_{k,\ell}| \leq B_{k,\ell } \leq \TOL(M).
	\end{equation}
Thus, the proposed strategy guarantees that the discarded SIs will be less or equal than $\TOL (M)$. Note that 
inequality (\ref{bound:TOL}) quantifies the error in the proposed methodology.

\vspace{5pt}
\noindent\textbf{Setting a pre-specified tolerance.}  On the other hand, a user  can also take advantage of
some pre-existing intuition regarding the modeled system, to argue that under a pre-specified tolerance the
SIs can be discarded as insensitive.
In this case we define the set,
\begin{equation}
C_{\TOL} := \left \{ \textrm{ all }    (k,\ell): \;  B_{k,\ell} \geq \TOL \; \right \},
\end{equation}
where $\TOL$ is the pre-specified tolerance.
As in the previous case, where the maximum number of SIs to be computed is fixed, the user will employ
a gradient estimation method (here a finite difference coupled algorithm is proposed, 
see eq. (\ref{derivative:estimator})) to compute the SIs, $S_{k,\ell}$, for all pairs $(k,\ell)$ in $C_{\TOL}$ with the 
guarantee that the discarded SIs will have magnitude less than $\TOL$.

\section*{Results}

In this section, we present and validate the proposed sensitivity analysis strategy in three
biological reaction networks. The first example is the p53 model which is a reaction network with
five reactions, three species and seven parameters. It is a small but interesting system due to the 
nontrivial long-time dynamics exhibiting random oscillations.
Here, p53 is used as an introductory example to present and test the proposed strategy. 
Then, we present and validate the proposed strategy for the
Epidermal Growth Factor Receptor
(EGFR) model  in the transient as well as in  the stationary regime showing that our method can be equally
applied at both regimes. Finally, we discuss a  protein homeostasis model  with a total number of
$4160$ SIs. This is a large-scale  realistic model with sloppy characteristics.

The  comparison  of computational costs between the proposed strategy and the  direct calculation of all SIs is discussed separately,
both in a general context as well as concretely  for the two latter examples; we refer to  
the  Discussion section below.


\subsection*{A p53 model}\label{section:p53}

\begin{figure}
	\centering
	\includegraphics[width=0.55\textwidth]{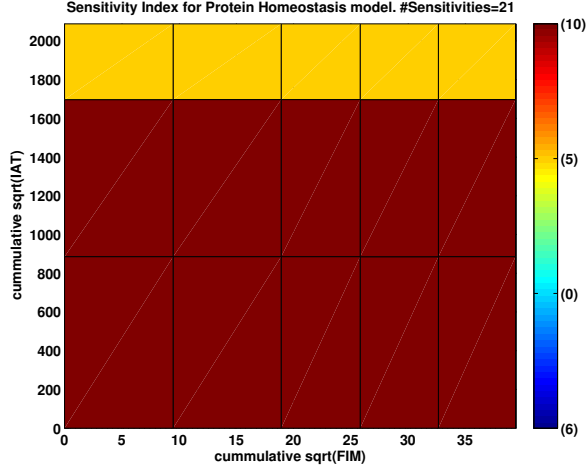}
	\caption{The cumulative square root of pathwise FIM (eq. (\ref{FIM:CTMC})) for the 7 parameters of p53 model is plotted
	versus the cumulative square root of the IAT (eq. \ref{IAT:CTMC}) for the three observables defined as the maximum
	amplitude of the PSD (eq. \VIZ{psd:max}). The area of the $(k,\ell)$ rectangle corresponds to the SB  of the SI, $S_{k,l}$.
	The red, yellow, light blue and dark blue rectangles correspond to values of the SB
	larger than 5000, between 5000 and 500, between 500 and 50 and less than 50,
	containing 10, 5, 0, and 6 SIs, respectively.}
	\label{fig:p53_strategy}
\end{figure}

The p53 gene  plays a crucial role for effective tumor
suppression in humans as its universal inactivation in cancer cells suggests \cite{Prives:98,
Harris:05, Geva-Zatorsky:06}. 
The p53 gene is activated in response to DNA damage and gives rise to  a negative feedback
loop with the oncogene protein Mdm2. Models of negative feedback are capable of oscillatory
behavior with a phase shift between the gene concentrations. Here, we validate the proposed sensitivity
analysis strategy to a simplified reaction network between three species,  p53, Mdm2-precursor and
Mdm2 introduced in \cite{Geva-Zatorsky:06}. The model  consists of five reactions and seven parameters
provided in Table~\ref{p53:reactions}. The nonlinear feedback regulator of p53 through Mdm2 
takes place in the second reaction while the remaining four reactions fall in the mass action
kinetics category.  Due to these mechanisms a nontrivial steady
state regime  characterized by random oscillations. 

\begin{table}[!htb]
\begin{center}
\caption{The reaction table where $x$ corresponds to p53, $y_0$ to Mdm2-precursor while $y$
corresponds to Mdm2. The state of the reaction model is defined as $\mathbf{x}=[y,y_0,x]^T$
while the parameter vector is defined as $\theta=[b_x,a_x,a_k,k,b_y,a_0,a_y]^T$.}
\begin{tabular}{|c|l|l|l|} \hline
Event & Reaction & Rate & Rate's derivative\\ \hline \hline
$R_1$ & $\emptyset \rightarrow x$ & $a_1(\mathbf{x}) = b_x$ & $\nabla_\theta a_1(\mathbf{x}) = [1,0,0,0,0,0,0]^T$ \\ \hline
$R_2$ & $x \rightarrow \emptyset$ & $a_2(\mathbf{x}) = a_x x + \frac{a_k y}{x+k} x$ & $\nabla_\theta a_2(\mathbf{x}) = [0,x,xy/(x+k),-a_k xy/(x+k)^2,0,0,0]^T$ \\ \hline
$R_3$ & $x \rightarrow x+y_0$ & $a_3(\mathbf{x}) = b_y x$ & $\nabla_\theta a_3(\mathbf{x}) = [0,0,0,0,x,0,0]^T$ \\ \hline
$R_4$ & $y_0 \rightarrow y$ & $a_4(\mathbf{x}) = a_0 y_0$ & $\nabla_\theta a_4(\mathbf{x}) = [0,0,0,0,0,y_0,0]^T$ \\ \hline
$R_5$ & $y \rightarrow \emptyset$ & $a_5(\mathbf{x}) = a_y y$ & $\nabla_\theta a_5(\mathbf{x}) = [0,0,0,0,0,0,y]^T$ \\ \hline
\end{tabular}
\label{p53:reactions}
\end{center}
\end{table}

Since the demonstrated model admits persistent, random oscillations we choose as observable the amplitude
of the oscillations 
for each of the three species. We extract the value of this observable from the Power Spectral Density
(PSD) \cite{Brown:Random:Signals} of the species time-series which is defined as the Fourier transform of the autocorrelation function
of the species time-series. The PSD of a continuous-time process, $X_t$, denoted by $|\hat X|^2$,
can be also given by
\begin{equation}\label{psd}
|\hat X|^2(\xi) = \frac{1}{T} \Big| \int_{0}^{T} X_{t} e^{i\xi t}  \;dt \Big|^2 \ .
\end{equation}
The maximum  amplitude of the PSD corresponds to the most prominent oscillation and it is given by
\begin{equation}
F(\{\mathbf{X}_t\}_{t=0}^T) = \max_{\xi} |\hat X|^2(\xi) \ .
\label{psd:max}
\end{equation}
This  observable is not in the form of (\ref{time:average}); however, our stationary sensitivity analysis as presented in (\ref{sens:bound:gen})--(\ref{FIM:CTMC}) still applies;
in order for the SB in (\ref{sens:bound:gen}) to be independent of the final time $T$ in the stationary regime, we have to prove
that the variance of $F$ scales like $O(\frac{1}{T})$. Indeed,
\begin{equation}
\begin{split}
	\VAR [ {F} (\{\mathbf{X}_t\}_{t=0}^T)] &= \EXPECT \big[ {F} (\{\mathbf{X}_t\}_{t=0}^T)^2\big] - \EXPECT \big[ {F} (\{\mathbf{X}_t\}_{t=0}^T)\big]^2  \\
				&\leq \frac{1}{T^2} \EXPECT \Bigl [ \int_0^T |X_t|  \; dt \Bigr],
\end{split}				
\end{equation}
showing that the variance of ${F}$ is of order $O(\frac{1}{T})$, provided $\EXPECT|\mathbf{X}_t|$ remains bounded.

In total, 21 SIs which correspond to 3 obrervables (max amplitude of PSD for each species
time-series) and 7 parameters must be computed. In Figure \ref{fig:p53_strategy}, the stationary
SB (eq. (\ref{sens:bound:statio})) of the SIs of the 3 observables with respect to the 7 parameters
is plotted as rectangle; the length of the x--side is equal to the square root of the pathwise FIM
(eq. (\ref{FIM:CTMC})) while the length of the y--side is equal to the square root of the IAT
(eq. (\ref{IAT:CTMC})). The area of the rectangle corresponds to the arithmetic value of the SB.
Thus, rectangles with small area indicate that the corresponding SI should also be small.
In Figure \ref{fig:p53_strategy}, the two least sensitive parameters have relatively so small
SB that cannot be distinguished in the plot. Notice that the grouping of the SIs of
Figure \ref{fig:p53_strategy} corresponds to the case ``Setting a pre-specified tolerance'' as described in 
the ``How to use the strategy'' section. Dictated by the average value of the observable functions
which take values in the range between $10^3$ and $10^4$, we set the tolerance values to 5000, 500 and 50.

\begin{figure}[htbp]
	\centering
	\includegraphics[width=0.99\textwidth]{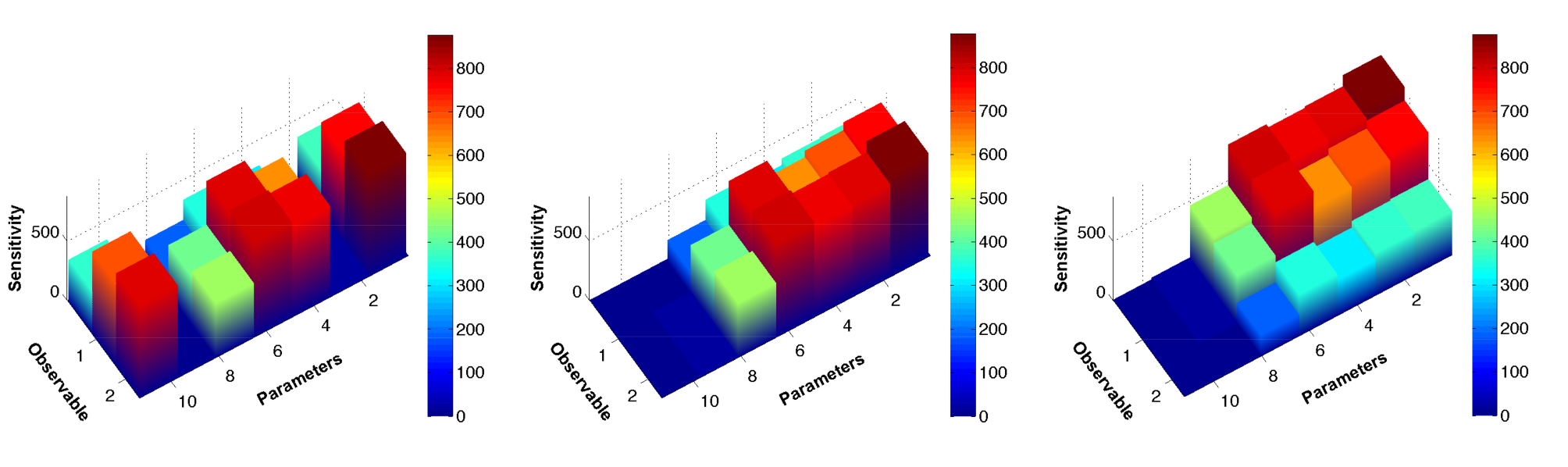}
	\caption{SIs for the maximum value of the PSD (eq. (\ref{psd:max})) of the three species of the p53
	model for $t\in[0,50]$ computed using the coupling gradient estimator (eq. (\ref{derivative:estimator})).
	No ordering of the observables or the parameters in the left plot.
	In the middle plot, the SIs are ordered in the parameter direction using the pathwise FIM
	(eq. (\ref{FIM:CTMC})). The ordering reveals that the pathwise FIM can serve as a first screening procedure to exclude
	the insensitive parameters. In the right plot, the estimated sensitivities are further ordered in the observable
	direction by sorting the IAT (eq. (\ref{IAT:CTMC})) in descending order.}
	\label{fig:p53_sensitivity}
\end{figure}

In order to validate the ordering of the SIs from high to low values based on the stationary SB,
we estimate them using the finite-difference coupling gradient estimator.
In the left plot of Figure \ref{fig:p53_sensitivity},  the SIs are plotted without any ordering, i.e., as they
are provided by the database \cite{url:biomodels}. In the middle plot, the SIs are ordered in
the parameter direction according only to the values of pathwise FIM. As expected from 
 Figure \ref{fig:p53_strategy} and the SB, the two least sensitive parameters have relatively very small SIs.
In the right plot of Figure \ref{fig:p53_sensitivity}, the SIs are further ordered according to the values
of IAT. The SIs with the largest values are concentrated to the right upper corner which correspond
to the larger SBs validating  {\em Step 1} of the proposed strategy.
Moreover, despite the fact that the SB correctly predicts the ordering of the SIs, the actual values of the SIs which
correspond to the SB that are labeled sensitive (red color in Figure~\ref{fig:p53_strategy}) differ by
an order of magnitude from the values of the SB making the {\em Step 2} of the proposed strategy
necessary for quantitative sensitivity results. On the other hand, the SIs that are labeled
as insensitive (dark blue color in Figure~\ref{fig:p53_strategy}) can be safely eliminated from the
{\em Step 2} since the SB are relatively close to zero (when compared to the remaining value of SB).

\begin{table}[!htb]
\centering
\caption{Parameter values for the p53 model.}
\begin{tabular}{|c||c|c|c|c|c|c|c|c|} \hline
Parameter & $b_x$ & $a_x$ & $a_k$ & $k$ & $b_y$ & $a_0$ & $a_y$ \\ \hline
Value & 90 & 0.002 & 1.7 & 0.01 & 1.1 & 0.8 & 0.8 \\ \hline
\end{tabular}
\label{p53:values}
\end{table}

\subsection*{An EGFR model}\label{sec:egfr}
The EGFR model is a well-studied reaction network describing signaling
phenomena of (mammalian) cells \cite{Moghal:99, Hackel:99,Schoeberl:02}. As its name suggests,
EGFR regulates cell growth, survival, proliferation and differentiation and plays a complex and crucial
role in embryonic development and in tumor progression \cite{Sibilia:98,Kim:99}. In this paper,
we study the reaction network developed by Kholodenko et al.\ \cite{Kholodenko:99} which consists
of 23 species and 47 reactions. 
\begin{figure}
	\centering
	\includegraphics[width=0.99\textwidth]{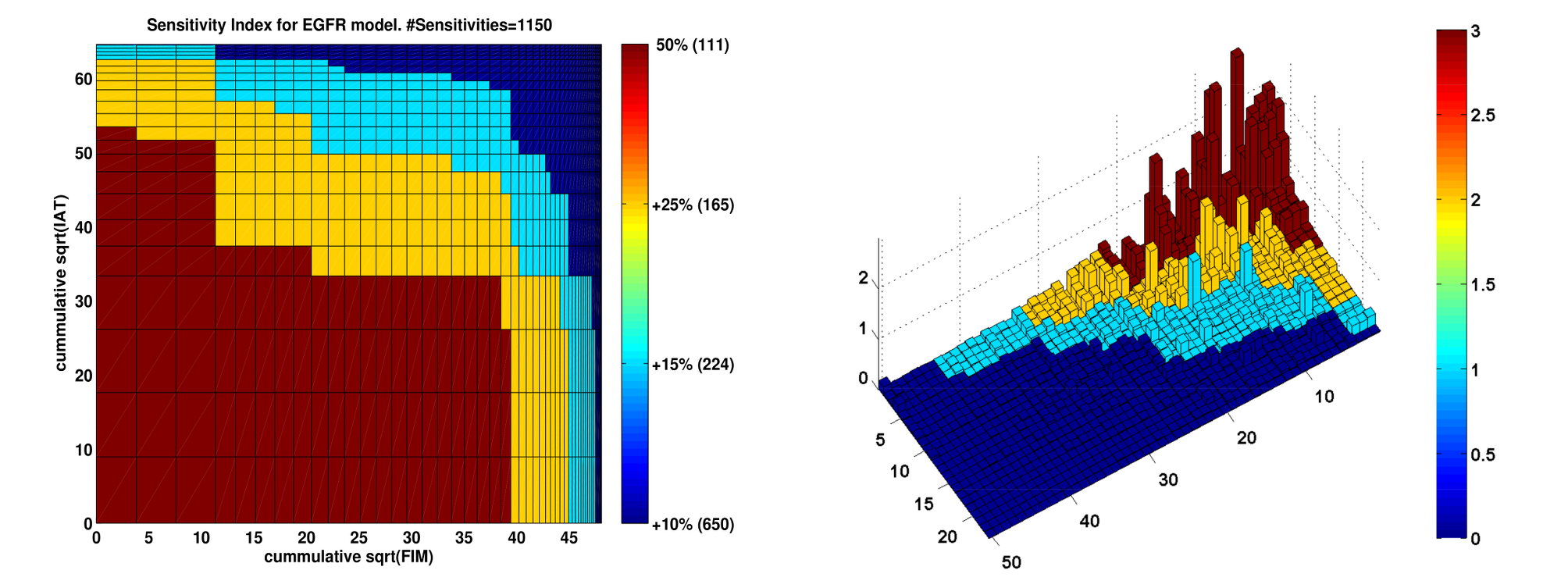}
	\caption{{\em Step 1} of the proposed sensitivity analysis strategy, graphically presented in the left plot,
	consists of identifying the least sensitive SIs. The area of  rectangle $(k,\ell)$ represents
	the SB of the SI, $S_{k,\ell}$, for the transient regime.
	Every color corresponds to a predefined percentage of the total area, e.g. blue and yellow
	correspond to 10\% and 25\% of the total area containing 650 and 165 SIs, respectively.
	The plot in the right represents {\em Step 2} of the strategy where the SIs
	are estimated by the coupling estimator (eq. (\ref{derivative:estimator})). The computed SIs
	are colored by the region identifier color of the left plot, showing that areas with low sensitivities
	correspond to areas with low values of the SB. }
	\label{fig:egfr_fim}
\end{figure}
The propensity function for the $R_{j}$ reaction of the EGFR network is written in the form (mass action kinetics, see \cite{Distefano:13})
\begin{equation}\label{mass:action}
a_j(\STATE) = k_j \binom{\STATE_{A_j}}{\alpha_j} \binom{\STATE_{B_j}}{\beta_j}, \quad  j=1,\ldots,47 \textrm{ and } j\neq 7,14,29 \ ,
\end{equation}
for a reaction of the general form ``$\alpha_j A_j + \beta_j B_j \xrightarrow{k_j} \ldots$'', where
$A_j$ and $B_j$ are the reactant species, $\alpha_j$ and $\beta_j$ are the respective number
of molecules needed for the reaction and $k_j$ the reaction constant. The binomial coefficient
is defined by $\binom{n}{k}=\frac{n!}{k!(n-k)!}$.  Here, $\STATE_{A_j}$ and $\STATE_{B_j}$ is
the total number of species $A_j$ and $B_j$, respectively. Reactions  $R_{7},R_{14},R_{29}$ are
exceptions with their propensity functions being described by the Michaelis--Menten kinetics, see \cite{Distefano:13},
\begin{equation}
a_j(\STATE) = V_j \STATE_{A_j} / \left(  K_j + \STATE_{A_j} \right), \quad j=7,14,29 \ ,
\end{equation}
where $V_j$ represents the maximum rate achieved by the system at maximum (saturating) substrate
concentrations while $K_j$ is the substrate concentration at which the reaction rate is half the maximum
value. The parameter vector contains all the reaction constants,
\begin{equation}
 \theta = [k_1,\ldots,k_6,k_8,\ldots,k_{13},k_{15},\ldots,k_{28},k_{30},\ldots,k_{47},V_7,K_7,V_{14},K_{14},V_{29},K_{29}]^T \ ,
\end{equation}
with $K=50$. In this study the values of the reaction constants are the same as in \cite{Kholodenko:99}.

\begin{figure}
	\centering
	\includegraphics[width=0.99\textwidth]{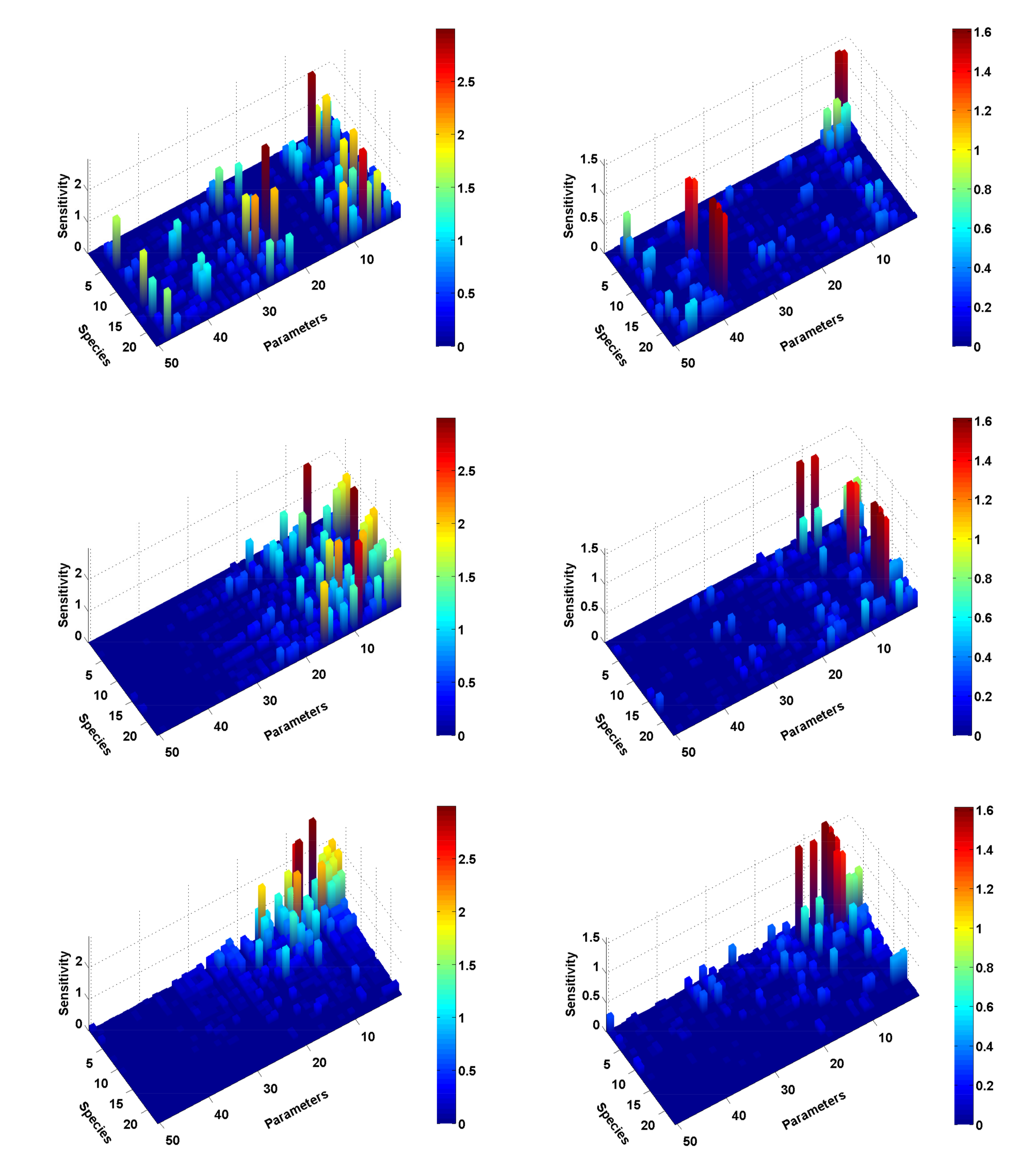}
	\caption{SIs for the EGFR model on the transient regime $t\in[0,50]$ (plots on left column)
	and on steady states regime $t\in[50,100]$ (plots on right column). The upper row presents
	the SIs without any ordering. On the middle row, the SIs are ordered
	in the parameter direction using the pathwise FIM (eq. (\ref{FIM:trans}) and (\ref{FIM:CTMC}) for the transient
	and stationary regime, respectively). On the lower row, the SIs are further ordered using the
	standard deviation of the observable and the IAT for the transient and stationary
	regime, respectively. }
	\label{fig:egfr_towers}
\end{figure}

Since the EGFR reaction network models signaling phenomena, it consists of a transient regime that corresponds to the
time interval $[0,50]$ and a stationary regime which approximately corresponds to the time interval $[50,\infty)$.
In this study, the computations in the steady states regime were done in the time interval $[50,100]$.
The general SB (eq. (\ref{sens:bound:gen})) and the stationary SB (eq. (\ref{sens:bound:statio})) are employed
for the transient and the stationary regime, respectively. Figure~\ref{fig:egfr_fim} presents results from
the transient regime.
In the left plot of Figure \ref{fig:egfr_fim}, the parameters (x--axis) and species populations (y--direction)
which define the observable functions are
sorted according to the square root of pathwise FIM and the square root of the variance of the observable,
respectively, as dictated by the SB. Thus, the area of every rectangle corresponds to the value of the
SB. This plot visualizes {\em Step 1} of the proposed strategy; the SI, $S_{k,l}$, corresponding to
a rectangle with small area can be safely excluded from {\em Step 2}
of the sensitivity analysis strategy. The coloring of the rectangles is performed as follows: starting from rectangles
with large area we color the first $50\%$ of the total area using red, the next $25\%$  using yellow and the
next $15\%$ and $10\%$ using light blue and dark blue, respectively.  This grouping is equivalent to
``Setting the maximum number of SIs'' as discussed in section ``How to use the proposed strategy''.

In the right plot of Figure \ref{fig:egfr_fim},  {\em Step 2} of the proposed strategy is depicted.
Even though all the SIs are computed using the coupling method for validation purposes, we
could exclude for instance the SIs that correspond dark blue area in the left plot of Figure \ref{fig:egfr_fim}
since these ``dark blue'' SIs are just a small portion of the total area of the rectangle. Thus,
approximately half of the SIs are excluded from {\em Step 2} and a upper bound (or tolerance)
is assigned to them.
Moreover, the right plot of Figure~\ref{fig:egfr_fim} serves as a validation of the proposed strategy; 
for instance, the SIs with large values are concentrated on the upper right corner (red and yellow in the
right plot of Figure~\ref{fig:egfr_fim}) which corresponds to the large values of the SB while the SIs
with small values (dark blue) are concentrated on the lower left corner validating the proposed strategy.

In Figure \ref{fig:egfr_towers}, the SIs for the transient regime (plots on left column) and the stationary regime
(plots on right column) are presented showing that the proposed strategy is capable of handling both regimes.
For further validation, all SIs are computed using the finite-difference coupling estimator (eq. (\ref{derivative:estimator})).
In the upper row of Figure \ref{fig:egfr_towers}, the SIs are unordered (arranged according to the database
ordering, see \cite{url:biomodels}). In the middle row of Figure \ref{fig:egfr_towers}, the SIs are ordered
in the parameter direction using only the pathwise FIM given by (\ref{FIM:trans}) for the transient
regime and by (\ref{FIM:CTMC}) for the stationary regime. Notice that this ordering produces also
a qualitative separation between insensitive and sensitive parameters. Hence, pathwise FIM alone can serve as an even  simpler
alternative to {\em Step 1} of the proposed strategy (see also the Discussion section, below). In the
lower row of Figure \ref{fig:egfr_towers}, the SIs are further ordered using the standard deviation of the
time-averaged observable and the IAT for the transient and stationary regime, respectively.
In both regimes, the SIs with large values are concentrated on the upper right corner
(lower row of Figure \ref{fig:egfr_towers}) which corresponds to the large values of the SB validating
the proposed strategy. Finally, notice that there are SIs for insensitive parameters (left side in lower row's plots)
with relatively non-negligible values, however they  stem from the statistical bias of the coupling method
and not from a wrong labelling of the SIs based on the SB.

\subsection*{A protein homeostasis model}
In \cite{Proctor:2011}, the authors propose a reaction network that models the role of two chaperones,
the Hsp70 and the Hsp90, in the maintenance of protein homeostasis. Loss of protein homeostasis is
the common link between many neuro-degeneration disorders which are characterized by the accumulation
of aggregated protein and neuronal cell death. The authors examined the role of both Hsp70 and Hsp90
under three different conditions: no stress, transient stress and high stress. Their model
was validated against experimental data.
The studied reaction network consists of $52$ species and $80$ reactions with propensities being of mass
action kinetics type described by (\ref{mass:action}). The reaction constants as well as the initial populations
were taken from \cite{Proctor:2011}. Defining as observables the averaged species populations, i.e.
$f_\ell(\bm{x})=x_\ell, \ l=1,\ldots,N$ in (\ref{time:average}),the total
number of SIs in this model is $4160$. The parameter vector consists of all reaction constants
\begin{equation}
\theta =  [ k_1,\ldots, k_{80} ]^T
\end{equation}
where the parameter values used here were taken from \cite{Proctor:2011}.

\begin{figure}
	\centering
	\includegraphics[width=0.99\textwidth]{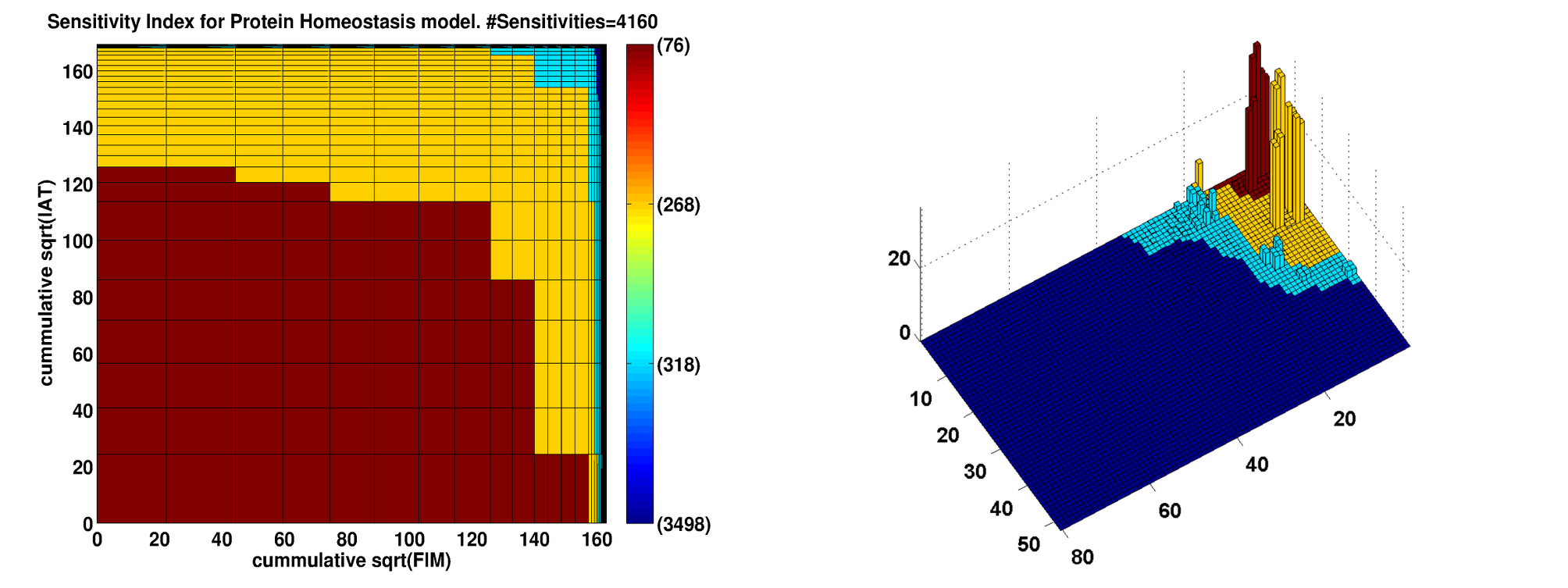}	
	\caption{{\em Step 1} of the proposed sensitivity analysis strategy consists of identifying the least sensitive
	parameters. In the left plot, the area of  rectangle $(k,\ell)$ is the SB
	of the SI, $S_{k,\ell}$. Red, yellow, light blue and dark blue  regions
	correspond to SB values greater than $100$, between $10$ and $100$, between $1$ and
	$10$ and  less than $1$, containing $76,268,318$ and $3498$ SIs, respectively. The
	plot in the right shows {\em Step 2} of the strategy and simultaneously serves as a validation of the proposed
	methodology. The actual SIs, computed by the coupling method,
	are colored by the region identifier color of the first plot, showing that areas with low sensitivities
	correspond to areas with low values of the SB.}
	\label{fig:protHom_strategy}
\end{figure}

In this model under the mechanism of no stress, only few SIs have large values while most of them
are close to zero presenting a good example for ``sloppiness" (see right plot of Figure~\ref{fig:protHom_strategy}).
Note that when more  complex mechanisms are included, the ``sloppiness" of the reaction network will
be changed but since we are interested in the validation of the proposed strategy,
we restrict our discussion 
in the no stress case. Due to the high number of SIs, it is of great importance to screen
out the insensitive pairs of parameters--observables using the stationary SB (eq.~(\ref{sens:bound:statio})).
Then, a more accurate and refined estimation of the potentially large SIs using the coupling method
can be performed ({\em Step 2} of the proposed strategy).


We group the SIs in the same way as in the p53 model, i.e.
the range of the estimated SB is divided into regions of the same order of
magnitude, e.g. the red and the dark blue region on the left plot of Figure \ref{fig:protHom_strategy}
correspond to SIs in which the SB has a value greater than 100 and less that 1, respectively.
The SIs with large values are concentrated on the upper right corner (right plot of
Figure~\ref{fig:protHom_strategy}) which corresponds to the large values of the SB
while the SIs with small values are concentrated on the lower left corner, validating once again
the proposed strategy. More precisely, {\em Step 1} of the proposed sensitivity strategy consists of
screening out the parameter--observable pairs that correspond to regions of rectangles with small area.
Assigning the value of the SB to the SIs with associated SB value less than 1 (dark blue in
Figure~\ref{fig:protHom_strategy}) results in a significant reduction in the total computational cost
since $3498$ out of $4160$ (approximately $85\%$) SIs can be safely discarded as insensitive.
Then, in {\em Step 2}, the SIs of the remaining pairs are computed using the coupling
method. As it can be seen in  the right plot of Figure~\ref{fig:protHom_strategy},
computing the SIs in red, yellow  and light blue areas is enough to obtain the important information
for the whole sensitivity matrix.

\section*{Discussion}
This section provides a detailed discussion on the computational gain of the proposed strategy as well as
a simple formula on the achieved speedup. Moreover, we discuss the error quantification in the proposed 
sensitivity analysis strategy.

\subsection*{Using (only) FIM to screen out insensitive parameters}
\begin{figure}
	\centering
	\includegraphics[width=0.99\textwidth]{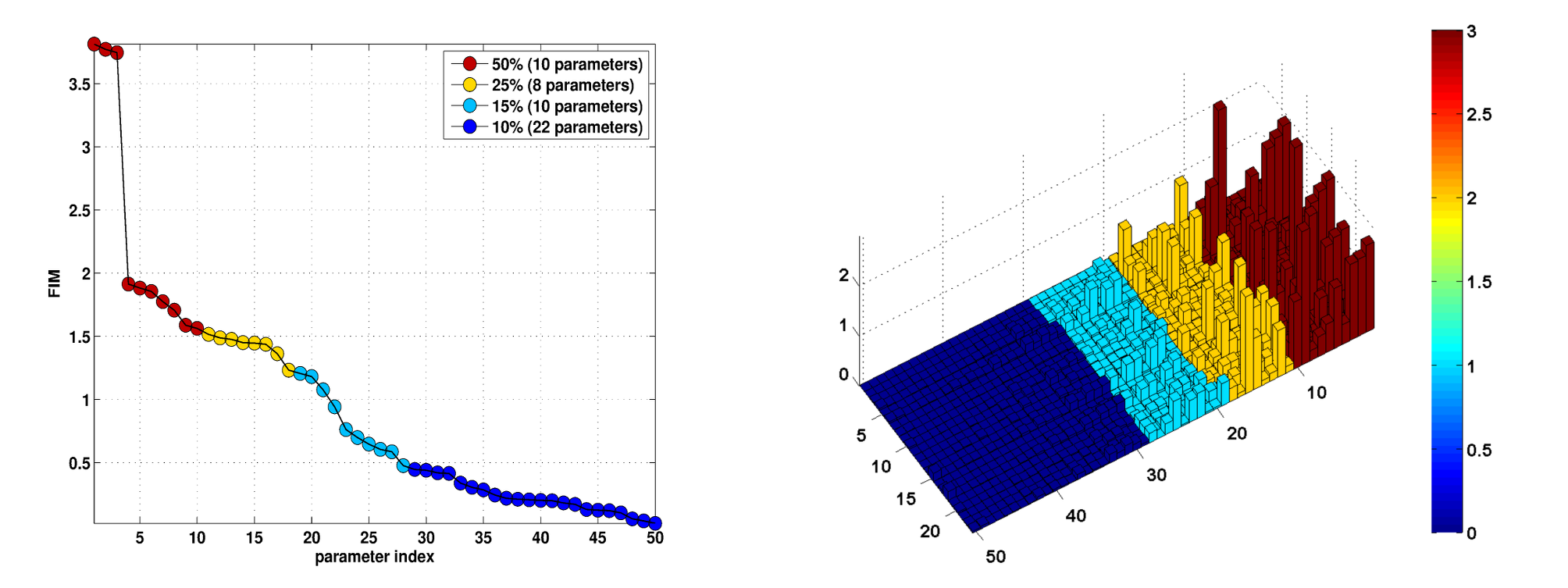}	
	\caption{ An alternative 1st step in the proposed strategy is presented in left plot where the pathwise FIM
	for the EGFR model is sorted in descending order. This ordering of the pathwise FIM gives a sorting in the parameters
	which gives a qualitative measure of the sensitivity of the model with respect to parameters.  In the second
	step of the strategy (right plot), the SIs computed using the finite-difference
	coupling estimator are ordered according to the pathwise FIM value of the left plot. This
	figure validates the fact that pathwise FIM alone can give a first qualitative estimate of SIs with respect
	to parameters.}
	\label{fig:EGFR:stepFIM}
\end{figure}

As discussed earlier (middle row of Figure \ref{fig:egfr_towers}) the pathwise FIM (eq. (\ref{FIM:trans}) or (\ref{FIM:CTMC}) in the steady state case)
can serve as a fast alternative screening method instead of the complete SB (inequalities (\ref{sens:bound:gen})
or (\ref{sens:bound:statio}) in the steady state case). In this case, the calculation
of the standard deviation of the observable (or the IAT (eq. (\ref{IAT:CTMC})) in the stationary case ) is
bypassed leading to a less accurate but  observable-independent screening method.  The ordering of the FIM from high
to low values provides a qualitative measure of sensitivity with respect to the parameters. 

In Figure \ref{fig:EGFR:stepFIM}, the pathwise FIM alone is utilized for the ordering
of the SIs for the EGFR model in the stationary regime, providing a computationally less expensive 
alternative to   {\em Step 1} of the strategy discussed earlier. In the left plot of Figure \ref{fig:EGFR:stepFIM},
the pathwise FIM is ordered in descending order. Then, the potentially most sensitive parameters
whose pathwise FIM values summing to $50\%$ of the total sum of the pathwise FIM are organized into the red group
and the following $25\%,15\%$ and $10\%$ are organized into three groups (yellow, light blue and dark blue, respectively).
Notice that this grouping is not unique and different parameter groupings can be used depending
on the model under consideration.
In the right plot of Figure \ref{fig:EGFR:stepFIM}, the SIs are computed using the coupling method and
sorted according to the ordering given by the pathwise FIM as in the middle row of Figure \ref{fig:egfr_towers}.
Moreover, the SIs are grouped and colored according to the grouping based on the pathwise FIM,
see left plot of Figure \ref{fig:EGFR:stepFIM}. It is evident that there
is a separation of the SIs into groups containing parameters with high (red and yellow), low
(light blue) and almost zero (dark blue) sensitivities. 
%

Although pathwise FIM as a sensitivity tool for reaction networks  was studied and used earlier in \cite{PKV:2013}, there are some
differences with the methodology proposed in this article. Here pathwise FIM serves as part of the upper bound of the
SI while in \cite{PKV:2013} there was no (immediate) connection of the pathwise FIM with the SIs.
Moreover, in \cite{PKV:2013} there was no estimation of the actual SIs while
here only the SIs of the most sensitive species/parameters  are estimated in {\em Step 2} of the proposed strategy. 
Nevertheless, estimates such as (\ref{sens:bound:gen}) and \VIZ{last} below give us quantified guarantees
to employ only the pathwise FIM, thus bypassing the costly {\em Step 2}. Of course, we  can only use this strategy  provided that
identifying and  screening out   insensitive parameters is the focus of  the sensitivity analysis. We also
refer to the ``Computational Cost" subsection below for  related  comments on this issue.

\subsection*{Error quantification in the accelerated sensitivity analysis}
In this section, we quantify the error of the proposed methodology
under the assumptions that only pathwise FIM is used in {\em Step 1} and the quantity of interest is $f_\ell(\mathbf{x}) = {x}_\ell,\ \ell=1,\ldots,N$, where $N$ is the number of the species.
We consider the case where there is a fixed amount of computational resources $M=K'\times N$,  where $K'$ is the maximum number of parameters in which SIs can be computed. 
Note that this is a special case of the error quantification discussed in section ``How to use the proposed strategy''.
Then the questions posed here are:
\begin{enumerate}
\item[(a)] For which parameters should the SIs  be computed?
\item[(b)] What is the magnitude of the SIs that are discarded by {\em Step 1} of the proposed strategy?
\end{enumerate}
To answer these questions we define the set 
\begin{equation}
C_{K'} := \left \{     \textrm{ all } k: k\textrm{-th diagonal element of pathwise FIM is one of the }  K' \textrm{ largest values}    \right \}
\end{equation}
Then, the answer to (a) is to compute SIs for all $k\in C_{K'}$. The answer to (b) is that for all $k\notin C_{K'}$ and $\ell=1,\ldots,N$, we have the bound
\begin{equation}
	|S_{k,\ell}| \leq \sqrt{ \tau_{\EQUIL}(f_\ell) } \max_{k\notin C_{K'}} \sqrt{\FISHERR(Q^\theta)_{k,k}}: = \TOL(\ell,K')\, ,
	\label{last}
\end{equation}
where the tolerance $\TOL(\ell,K')$ is defined as above and the last inequality follows from the stationary SB (ineq. \VIZ{sens:bound:statio}).
The SB given by \VIZ{last}  assures that the error in the proposed methodology due to the discarded (by {\em Step 1})
SIs, will be less or equal to $\max_{\ell}\TOL(\ell,K')$.

\subsection*{Computational Cost}
The computational cost of the proposed strategy, consists of the cost of the estimation of the SB (ineq. (\ref{sens:bound:gen})
or (\ref{sens:bound:statio})) as well as the cost of estimation of the SIs not discarded from {\em Step 1} by 
using the coupling method. This cost is compared with the computational cost of using the coupling method
for all SIs without the screening step of {\em Step 1}.
The comparison will be done in terms of  the computational gain $G$, which is the sum of the costs from
{\em Step 1} and {\em Step 2} of the strategy over the cost of the coupling method for all SIs:
\begin{align}\label{gain:in:words}
G &:= \frac{\textrm{cost of proposed strategy}}{\textrm{cost of computing all SIs}}  \nonumber \\
&= \frac{\textrm{cost of SB}}{\textrm{cost of computing all SIs}} + \frac{\textrm{cost of computing sensitive SIs}}{\textrm{cost of computing all SIs}} \nonumber \\
&= G_1 + G_2 \ .
\end{align}
Note that  1/G  measures how many times the proposed strategy is faster compared to the estimation of all SIs using the coupling method.
 For the technical details on this comparison see File S4 in the supporting information. 

{ For simplicity we consider the case where only the pathwise FIM is used to discard insensitive parameters, see the previous section
``Error quantification in the accelerated sensitivity analysis". } In this case  information on the sensitivity of observables is obtained from \VIZ{last} using 
{\em Step 1}. 
Moreover,  the 
quantity of interest (observables) considered here  is the species population, i.e. $f_\ell(\mathbf{x}) = {x}_\ell$.  Finally, the comparison is being done under the 
requirement that the \textit{relative confidence intervals} (for the definition,
see File S4 ) of all estimators involved in both approaches will have variance less or equal to $\delta\ll 1$.

We observe that the variance of the SB is much less than the variance of the
coupling method (i.e. eq. (\ref{var:estimator})) leading to $G_1 \ll 1$ (see Figure 1 in File S4).
After the computation of the SB, the SIs are grouped into two categories, i.e.,
potentially sensitive and insensitive. Let $K'$ be the number of sensitive parameters that correspond to the potentially sensitive SIs
and $K$ the number of all
parameters. Then, under some reasonable assumptions (see File S4 for more details), the
$G_2$ term is approximately equal to $G_2 = \frac{K'}{K}$ and the computational gain $G$ is approximated by
\begin{equation}\label{gain:approx}
G \approx \frac{K'}{K}.
\end{equation}

We next validate the approximation (\ref{gain:approx}) of the computational gain on the EGFR model
presented earlier, where the number of parameters is $K=50$. By inspection of Figure \ref{fig:EGFR:stepFIM} the
number of sensitive parameters is $K'=28$, which correspond to the first three  colored regions of the graph.
Thus, $G$ can be approximated by $G\approx\frac{28}{50}=0.56$.
On the other hand, we compute the $G_1$ and $G_2$ terms exactly by measuring the
simulation cost of the two methods in terms of counting the number of required samples. The $G_1$ term is
equal to $0.0034$ showing that the estimation of the SB needs about $300$ times less samples than
that of computation of SIs using the coupling estimator while $G_2$ is equal to $0.72$. As a result, the actual speed-up due to the proposed strategy is approximately 1/G$\approx 1.4$. These are modest computational gains, however they are achieved with minimal investment in computational resources in  {\em Step 1\,};  on one hand,  the variance calculation in the SB is anyhow necessary in any forward simulation,  in order to obtain confidence intervals for the  species (observables), while the calculation of the pathwise FIM is straightforward and can be viewed as the simulation of just one  additional observable.

\begin{figure}
	\centering
	\includegraphics[width=0.99\textwidth]{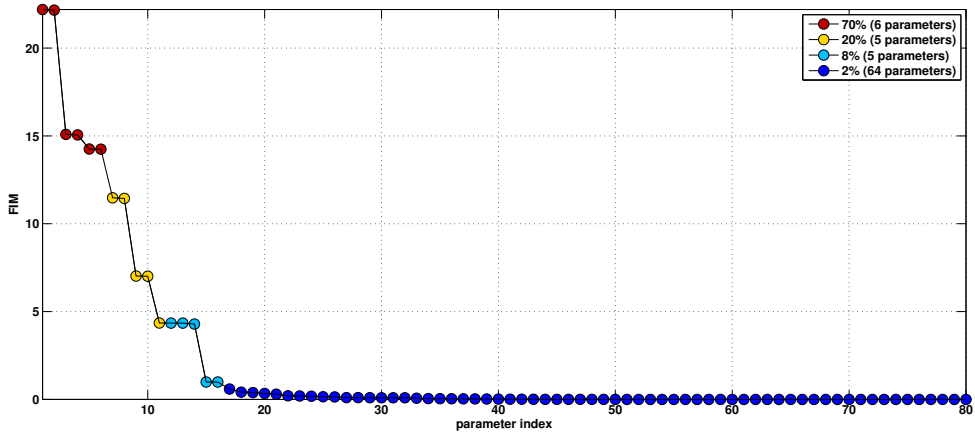}
	\caption{Pathwise FIM for the Protein Homeostasis model sorted in descending order. The coloring of the parameters is being done as follows: parameters whose pathwise FIM values sum to 70\%, 20\% ,8\% and 2\% of the total sum of pathwise FIM values, are grouped together and correspond to red, yellow, light blue and dark blue color, respectively.}
	\label{fig:protHom:FIM}
\end{figure}

Although the computational gains in systems such as EGFR,  where a large number of parameters are relatively sensitive are modest, the gains are very significant in ``sloppy" systems.
Indeed, for the Protein Homeostasis reaction network, assuming that $G_1$ is negligible compared to $G_2$
and using the approximation in (\ref{gain:approx}) to obtain an estimate for $G_2$, we have that the computational gain
for this model is $G\approx \frac{K'}{K} = \frac{16}{80} = 0.2$.  The value for $K'$ is obtained by assuming
that the important parameters are those colored with red, yellow and light blue in  Figure \ref{fig:protHom:FIM}.
The value of $G\approx 0.2$ suggests a 5-times speed-up in the sensitivity analysis of this "sloppy" example.

Finally, as discussed  at the end of  the subsection ``Using (only) FIM to screen out insensitive parameters", we may also   employ just {\em Step 1} (and skip {\em Step 2}), at least when  
 identifying and  screening out   insensitive parameters is the focus of  the sensitivity analysis. This may be the case in high-dimensional reaction networks  with suspected ``sloppy" characteristics. In this case, the computational cost of the proposed methodology dramatically decreases.  For instance, in the EFGR case the speed-up is 1/$G_1 \approx$ 294 times faster than the full coupling method. In the case of the Protein Homeostasis example the gains are much higher.

\section*{Conclusions}

Existing information-based parametric sensitivity analysis methods for stochastic reaction networks can  tackle
systems with a large number of parameters without however providing insights on specific   quantities of interest (observables).
 Furthermore, existing methods can perform accurate
sensitivity analysis  for arbitrary observables but with high computational cost which can  become prohibitive  for networks with a high dimensional  state and/or parameter  space 
due to (a) the high variance  of SI estimators, and/or (b)  the need to calculate  gradients corresponding to all parameters.
In the proposed methodology, we address these challenges  through a two-step strategy, combining  two different sensitivity analysis approaches using a new 
SI upper bound. More specifically, in {\em Step 1}, we first perform  an
``insensitivity'' analysis: namely,  the parameters' sensitivity  can be  systematically screened and many can be eliminated as insensitive  based on derived 
SBs of the SI which are computationally inexpensive; in {\em Step 2}, only   the potentially sensitive parameters which were not screened out in {\em Step 1} are  estimated exactly, based on the finite-difference (gradient)  
coupling approach.

The acceleration in sensitivity analysis due to  the proposed strategy can be very significant  especially when sloppy systems
are considered  and most of the parameters are expected to  be screened out as insensitive from {\em Step 1}. Moreover, the
proposed strategy offers a simple way to rationally balance  accuracy and computational cost, selecting 
the number of  insensitive parameters that need to be discarded. Specifically, the tradeoff between computationally expensive  gradient
estimation and accuracy in SI computation is quantified in terms of an easily computable SBs on the SIs
(see \VIZ{sens:bound:gen} and \VIZ{sens:bound:statio} for the transient and the stationary regimes, respectively), which in turn determines a cutoff (or a
user-determined tolerance) of insensitivity. The proposed strategy, through  the SB, guarantees that
the SIs for the insensitive parameters will lie below the value of the cutoff.
Thus, it is upon the practitioner's choice how many of the parameters will be screened out, based on the SB values and the overall computational budget.
 The computational acceleration of the
proposed strategy is approximately quantified by the ratio between the total number of parameters
over the number of the potentially sensitive parameters which were not eliminated in {\em Step 1}, i.e., the ratio $\frac{K}{K'}$.

We conclude this paper by noting  that the proposed strategy is by no means restricted to well-mixed
systems such as reaction networks and  can be directly applied  to spatially-extended systems (high-dimensional in state space). Indeed, the pathwise FIM
used in the screening in {\em Step 1},  still has low variance for spatially-extended systems such as Kinetic Monte Carlo, as it has been shown in catalysis examples 
\cite{Pantazis:Kats:13}, while the coupling method  in {\em Step 2} can be modified for such models so that it still gives reduced-variance
estimators for the SIs, see  \cite{AK:2013}. In fact, the proposed strategy is absolutely necessary in
spatially-expended systems with a large number of parameters since gradient computations are prohibitively expensive due to the need
for  a large number of repeated runs arising in the computation of SIs. Finally, the proposed strategy is
compatible with any other sensitivity analysis approach in the sense that any gradient estimation
method can be utilized in {\em Step 2} instead of the coupling method.

\section*{Supporting Information}
\noindent\textbf{File S1. Information Theory and Sensitivity Bounds.} The SBs are presented  from an
information theory perspective. The general SBs (both transient and
stationary) are obtained by a limiting process on the relative entropy between path distributions.
The information theory perspective provides also intuitive and explicit formulas for the quantities
of interest for both discrete-time Markov chains and continuous-time Markov chains.

\vspace{5pt}
\noindent\textbf{File S2. Unbiased Statistical Estimators for pathwise FIM and IAT.} 

\vspace{5pt}
\noindent\textbf{File S3. Coupling of Stochastic Processes.} A brief revision of the coupling method proposed
by Anderson for the estimation of gradients in well-mixed reaction network systems.

\vspace{5pt}
\noindent\textbf{File S4. Computation Cost based on Variance Estimates.} An approximation of the variance of the SB
is first presented. Then, we present the technical details for the computational cost comparison of the
proposed strategy with the coupling method applied to all the SIs of the model.

\vspace{5pt}
\noindent\textbf{File S5. Matlab code.}  A zip file that contains all the Matlab source files needed for
the generation of the figures of this publication.


\bibliography{strategies}

\end{document}